\documentclass[12pt]{article}
\usepackage[all]{xy}

\usepackage{etex}
\usepackage{amsmath,amssymb,amscd,bm,mathtools,array}

\usepackage[margin=1.75cm]{caption} 
\usepackage[labelfont=it,textfont={small,it}]{caption} 

\usepackage{color}
\usepackage[colorlinks=true, pdfstartview=FitV, linkcolor=blue, citecolor=blue, urlcolor=blue]{hyperref}

\usepackage[cyr]{aeguill}     
\usepackage[pdftex]{graphicx} 
\DeclareGraphicsExtensions{.jpg,.mps,.pdf,.png} 

\let\ssection=\section
\renewcommand{\section}{\setcounter{equation}{0}\ssection}

\newcommand{\la}{{\langle}}
\newcommand{\ra}{{\rangle}}

\newcommand{\bone}{{\bf 1}}

\newcommand{\bgamma}{\boldsymbol{\gamma}}

\newcommand{\const}{{\mathrm{const}}}

\newcommand{\cE}{{\mathcal{E}}}

\newcommand{\rE}{\mathrm{E}}

\newcommand{\rg}{\mathrm{g}}

\newcommand{\cL}{{\mathcal{L}}}

\newcommand{\bcL}{\bm{\cL}}

\newcommand{\rO}{\mathrm{O}}

\newcommand{\so}{\mathfrak{so}}

\newcommand{\bomega}{\boldsymbol{\omega}}

\newcommand{\barP}{\overline{P}}
\newcommand{\bbP}{\mathbb{P}}

\newcommand{\bp}{{\mathbf{p}}}
\newcommand{\hbp}{\bp}
\newcommand{\hbpe}{\bp_e}

\newcommand{\nhpe}{\Vert\hbpe\Vert}
\newcommand{\nhp}{\Vert\hbp\Vert}
\newcommand{\Pf}{\mathrm{Pf}}
\newcommand{\dP}{\dot{P}}
\newcommand{\cP}{{\mathcal{P}}}
\newcommand{\bcP}{\bm{\cP}}

\newcommand{\barQ}{\overline{Q}}

\newcommand{\bbR}{\mathbb{R}}

\newcommand{\Ric}{\mathrm{Ric}}

\newcommand{\bs}{{\mathbf{s}}}
\newcommand{\ns}{\Vert\bs\Vert}

\newcommand{\sign}{\mathrm{sign}}

\newcommand{\SU}{\mathrm{SU}}

\newcommand{\dS}{\dot{S}}

\newcommand{\Tr}{\mathrm{Tr}}

\newcommand{\cT}{{\mathcal{T}}}
\newcommand{\bbT}{\mathbb{T}}

\newcommand{\hW}{W}

\newcommand{\dX}{\dot{X}}
\newcommand{\bx}{{\mathbf{x}}}

\newcommand{\bz}{{\mathbf{z}}}

\newcommand{\half}{\frac{1}{2}}

\newcommand{\de}{d}

\newcommand{\lb}{\left[}
\newcommand{\rb}{\right]}
\newcommand{\lp}{\left(}
\newcommand{\rp}{\right)}
\newcommand{\bb}{\begin{eqnarray}}
\newcommand{\ee}{\end{eqnarray}}
\newcommand{\eee}{\nonumber\end{eqnarray}}
\newcommand{\qq}{\quad}
\newcommand{\dpp}{\vcentcolon}

\def\s{\sigma}

\title{
Gravitational birefringence of light\\
in Robertson-Walker cosmologies
}

\author{
C. Duval\footnote{mailto: duval@cpt.univ-mrs.fr}
\ \ and\ \ 
T. Sch\"ucker\footnote{mailto: schucker@cpt.univ-mrs.fr}
\\[8pt]
Aix Marseille Univ, Universit\'e de Toulon, CNRS, CPT,
Marseille, France
}

\begin{document}

\date{August 29, 2017}

\maketitle

\thispagestyle{empty}

\abstract{
The spacetime evolution of massless spinning particles in a Robertson-Walker background is derived using the deter\-ministic system of equations of motion due to Papapetrou, Souriau and Saturnini. 
A numerical integration of this system of differential equations in the case of the standard model of cosmology is performed. The deviation of the photon worldlines from the null geodesics is of the order of the wavelength. Perturbative solutions are also worked out in a more general case. An experimental measurement of this devia\-tion would test the acceleration of our expanding universe.
}

\vspace{1cm}

\hfill {\em To the memory of Pierre Bin\'etruy}

\vspace{1cm}

%
%
%
%
%


\baselineskip=19pt

\section{Introduction}

\begin{figure}[h]
\begin{center}
\includegraphics[scale=0.99]{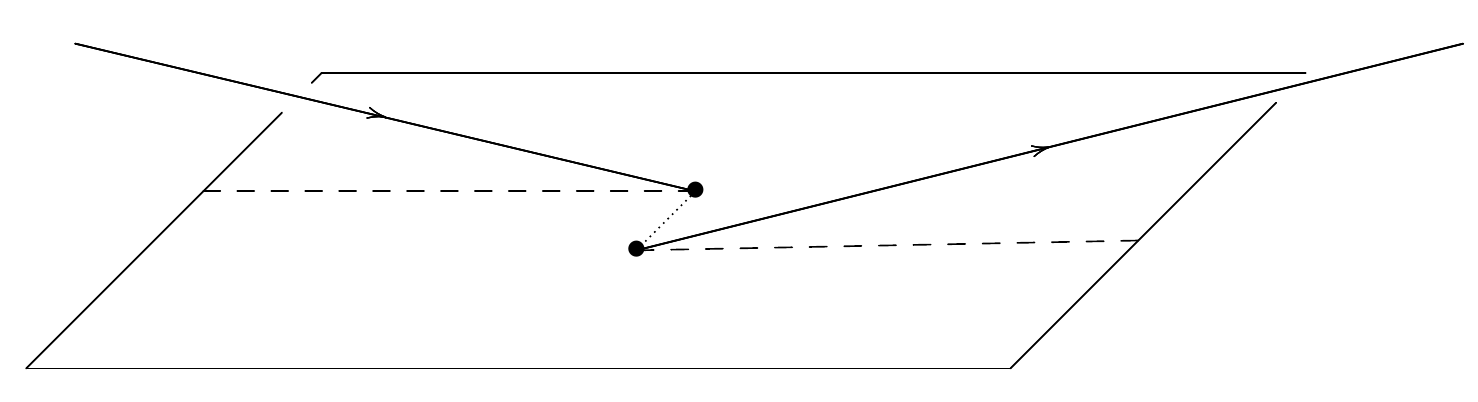}
\caption {The Fedorov-Imbert effect for reflection: A plane glass surface (represented by the rectangle) reflects an incoming, circularly  polarized  light beam. The dashed lines indicate the orthogonal projections of incoming and reflected light beams onto the glass surface. The dotted line (between the blobs) is the offset between the reflected beam and a hypothetical reflected beam of spinless photons (not shown in the figure). The offset is of the order of the wavelength of the light beam.
\label{fi}}
\vspace{-.8cm}
\end{center}
\end{figure}

The discovery of the effect of photonic spin on the trajectory of polarized light rays (energy fluxes) in optical media should historically be credited to Fedorov \cite{Fed55} and Imbert \cite{Imb72} (for reflection).  This subtle deviation from the Snell-Descartes law, see Figure \ref{fi} has been recently revisited and is today known under the name of Spin Hall Effect of Light (SHEL) either for reflection or for refraction. As shown in \cite{BB04,OMN04,BB06} by means of a semi-classical limit of Maxwellian wave optics, light rays experience a tiny shift depending upon their polarization state. This offset is of the order of magnitude of the wavelength and sideways, namely normal to the incidence plane on an interface. The effect (not to be confused with the longitudinal Goos-H\"anchen effect interpreted in terms of evanescent waves) can be understood as a sort of photonic Magnus effect \cite{BB04}. It is described in terms of a Berry connection introduced in order to account for the interaction of the spin with the ``gradient'' of the physical field, namely the refractive index in this case. Quite a large number of articles following these references have, since then, been published in this rapidly evolving subject in optics; see, e.g., \cite{BA13} for an up-to-date overview. At this stage, it should be emphasized that the SHEL, originally studied from a theoretical perspective, has lately been observed experimentally using techniques of Weak Quantum Measurement~\cite{BNKH08,HK08} that are well adapted to wavelengths in the nano\-meter range. Hence the subject rests on strong theoretical and also experimental bases.


It is noteworthy that the SHEL was shown to admit, rather unexpectedly, a full-fledged  description \cite{DHH1,DHH2} in terms of symplectic geometry based on the generic coadjoint orbits of the Euclidean group $\rE(3)$ with ``built-in'' Berry connection. This formalism was then used to derive the equations of motion of photons in arbitrary inhomogeneous \cite{DHH1}, anisotropic~\cite{Duv08} optical media, as well as polarized classical light rays in inhomogeneous media \cite{Duv12}. The crux of the theory was the  occurrence, via plain gravitational minimal coupling, of a spin-curvature coupling term responsible for an anomalous velocity. It is this specific geometrical standpoint, conveniently adapted to general relativity (GR), that we will espouse in the present work.

With the advantage of our previous experience with SHEL, our purpose will therefore be two-fold. We will first set up a purely geometric (and classical) formalism to describe the motion of spinning massless particles in GR. They are governed by highly non-linear ordinary differential equations presented in Section \ref{SouriauSaturniniSection} and specialized to the setting of Robertson-Walker universes subsequently. Secondly, we will integrate numerically these equations in the flat Lambda Cold Dark Matter model to analyse the new spin-effects. They again feature offsets of the worldlines of light from the conventional null geodesics used to define the redshift in cosmology. Again the offsets are of the order of magnitude of the wavelength. 

Let us mention, for completeness, independent contributions to the description of spinning light rays in GR, starting from semi-classical approximations of the Maxwell equations in a stationary \cite{FS11} and static gravitational background \cite{GBM07}. For an overview of the subject, let us refer to the review \cite{BN15}, and the textbook \cite{TB15}. This is however not the approach that we will follow.

It is also worthwhile recalling that the search for a consistent set of differential equations governing the motion of classical spinning particles in a general relativistic background has been a long-term endeavor for physicists. It certainly goes back to the first breakthrough by  Mathisson \cite{Mat37} and Papapetrou~\cite{Papa51}. See also the more recent contri\-butions \cite{Tau64,Dix70,DFS72,Sou74,Ste78,ZHA14}, and \cite{BI75} for relevant references. It would be inconceivable to provide here a full and reasonably complete bibliography for this well-studied subject which has, since then, triggered new developments in symplectic geometry, super\-geometry, etc.

Now, the main objective of the previous references was to describe the motions of massive spinning particles in GR. Curiously enough, the case of massless particles endowed with a classical spin seems to have escaped full consideration; see, however, Saturnini's thesis~\cite{Sat76} in the wake of Souriau's foundational memoir \cite{Sou74}. We also refer to the more recent contribution \cite{EDHZ16}.

The article is organized as follows.

In Section \ref{PSS} we introduce the universal Mathisson-Papapetrou-Dixon Souriau equations of motion of classical spinning particles in an external gravitational field. We choose to take advantage of the geometric derivation proposed by Souriau \cite{Sou74} based on a noteworthy formulation of the principle of general covariance. The model is then turned into a deterministic one by means of the so-called phenomenological Tulczyjew condition expres\-sing monolocality of the system. The massless case follows in a straightforward manner and leads to a new set of equations of motion credited to Souriau \cite{Sou74} and Saturnini \cite{Sat76}. The latter system of differential equations is at the root of our analysis of the trajectories of spinning photons in the standard cosmological model. It is independently derived in Appendix \ref{AppendixA}. Moreover, universal Noetherian conservation laws associated with spacetime isometries (and conformal rescalings) are worked out in a geometrical way.

\goodbreak

Section \ref{PhotonsRWSection} sets up the explicit form of the equations of motion of massless spinning photons with helicity $s=\pm\hbar$ in a general Robertson-Walker background. These equations turn out to describe the time evolution $t\mapsto(\bx(t),\hbp(t),\bs(t))$ of the position in comoving coordinates, the spatial momentum, and the spin vector. Let us emphasize that spin is no longer parallel to the momentum in GR under a mild assumption concerning the spin-coupling term. The spin of photons is therefore not ``enslaved'' \cite{DH15} and the resulting extra degrees of freedom are responsible for their localization in curved spacetime --- as opposed to their well-known delocalization in  Minkowski spacetime (see Footnote \ref{FootnoteDeleocalization}). This is the fundamental point exploited here to describe the offset of the photon worldlines from standard null geodesics.

Section \ref{NumericalSolutionsSection}  presents a numerical solution to the equations of motion in the particular case of the flat Lambda Cold Dark Matter model.

Section \ref{PerturbativeSolutionsSection} returns to flat Robertson-Walker universes and presents the general perturbative solution to the equations of motion in first order.

\section{Spinning light rays in general relativity}\label{PSS}

Let us first recall how the equations governing the motion of spinning classical particles arise in general relativity. 
To that end, we shortly review Souriau's derivation \cite{Sou74} based on the principle of general covariance applied to the description of matter distributions.
We will then show how these equations of motion specialize to the massless case once a specific equation of state has been imposed.

\subsection{The Mathisson-Papapetrou-Dixon equations}

Matter configurations may be viewed as distributions on the set of all Lorentzian metrics, $\rg$, of a given spacetime manifold, $M$. (We will choose the signature $(-,-,-,+)$, and assume $M$ to be time and spacetime oriented.) In particular, localized spinning particles will be described by first-order tensor distributions, $\cT_C$, supported by a world\-line~$C$, parametrized by $\tau$. Those are of the general form  
\begin{equation}
\la\cT_C,\delta{\rg}\ra=\half\int_{C}{\left[T^{\mu\nu}\,\delta{\rg}_{\mu\nu}+\Theta^{\mu\nu\rho}\,\nabla_\mu\delta{\rg}_{\nu\rho}\right]d\tau}
\label{TC}
\end{equation}
where the compactly supported variations, $\delta g$, of the metric serve as test-functions; here, $T\,d\tau$ and $\Theta\,d\tau$ are tensor densities of the  curve $C$ that define the distribution~$\cT_C$.\footnote{Greek letters $\mu,\nu,\rho,\ldots$ represent everywhere spacetime coordinate indices varying from $1$ to $4$.} 

\goodbreak

According to Souriau, the principle of general covariance does apply in this framework by assuming that the value $\la\cT_C,\delta\rg\ra$ be actually \textit{independent} of the metric in the orbit of~$\rg$ under all compactly supported diffeomorphisms of $M$; this translates infinitesimally to the following invariance property, namely
\begin{equation}
\la\cT_C,L_Y\rg\ra=0
\label{TCLYg=0}
\end{equation}
for all vector fields, $Y$, with compact support in $M$. In (\ref{TC}) and (\ref{TCLYg=0}), $\nabla$ and $L$ represent the Levi-Civita connection and the Lie derivative, respectively.

This point of view, due to Souriau \cite{Sou74}, is extremely convenient for the following purposes:
\begin{itemize}
\item
It solves the problem of the equations of motion, using only the invariance under the group of spacetime diffeomorphisms of an otherwise unspecified matter Lagrangian.
\item
In the spinless case, it provides an elegant derivation of the geodesic equations, both in the massive and massless case.
\item
It also works for extended particles, namely strings, branes, etc. In the case of a continuous matter distribution, it leads readily to the conservation of the energy-momentum tensor.
\item
For pointlike spinning particles, it yields in a straightforward manner the universal Mathisson-Papa\-petrou-Dixon (MPD) equations.
\item
It produces in an algorithmic fashion the Noetherian conservation laws associated with Killing vector fields independently of the choice of a matter Lagrangian. 
\end{itemize}
The last two purposes will be what we are mainly concerned with in this article.

Elaborate calculations \cite{Sou74} show that (\ref{TCLYg=0}) implies the following form of (\ref{TC}), namely
\begin{equation}
\la\cT_C,\delta{\rg}\ra=\half\int_{C}{\left[P^\mu\dot{X}^\nu\,\delta{\rg}_{\mu\nu}+S^{\mu\nu}\dot{X}^\rho\nabla_\mu\delta{\rg}_{\nu\rho}\right]d\tau}
\label{TCbis}
\end{equation}
where the linear momentum $P=(P^\mu)$, the velocity $\dX\dpp=dX/d\tau=(\dX^\mu)$, and the spin tensor $(S^{\mu\nu})$, verifying $S^{\mu\nu}+S^{\nu\mu}=0$, at the spacetime location $X(\tau)\in{}C$, turn out to satisfy the so-called \textit{universal Mathisson-Papa\-petrou-Dixon (MPD) equations} \cite{Mat37,Papa51,Dix70}
\begin{eqnarray}
\label{dotPbis}
\dP^\mu&=&-\half{}{R^\mu}_{\rho \alpha\beta}S^{\alpha\beta}\dX^\rho\\
\label{dotSbis}
\dS^{\mu\nu}&=&P^\mu\dX^\nu-P^\nu\dX^\mu.
\end{eqnarray}
Here 
${R^\mu}_{\nu\alpha\beta}=\partial_{\alpha}{\Gamma^\mu}_{\beta\nu}-\partial_{\beta}{\Gamma^\mu}_{\alpha\nu}+\cdots$
denote the components of the Riemann curvature of the Levi-Civita connection of $(M,\rg)$, and the dot stands for the covariant derivative along~$C$, with respect to $\tau$.

In some parts of the calculation we use linear maps, e.g., $S=({S^\mu}_\nu)$, rather than $2$-tensors, $(S^{\mu \nu})$, where ${S^\mu}_\nu\dpp=S^{\mu \rho }g_{\rho \nu}$. We will refer to the former when we suppress indices, e.g., $\Tr(S^2)\dpp={S^\mu}_\nu{S^\nu}_\mu$. The skewsymmetry of $(S^{\mu \nu})$ translates to the skewsymmetry of~$S$ with respect to the metric: 
$\rg(SV,W)=-\rg(V,SW)$ for all vectors $V$ and $W$.
In the same vein we denote by $\barP$ the covector associated to the vector $P$ via the metric: $\barP_\mu \dpp=g_{\mu \rho }P^\rho $.

The equations (\ref{dotPbis}) and (\ref{dotSbis}) are widely accepted as those governing the motions of classical spinning particles under the influence of a gravitational field. They thus read in coordinate-free form
\begin{eqnarray}
\label{dotP}
\dP&=&-\half{}R(S)\dX\\[6pt]
\label{dotS}
\dS&=&P\overline{\dX}-\dX\barP
\end{eqnarray}
with the skewsymmetric linear map $R(S)$ defined by
\begin{equation}
{R(S)^\mu}_\nu={R^\mu}_{\nu\alpha\beta}\,S^{\alpha\beta}.
\label{R(S)}
\end{equation}

In the sequel, we will use the shorthand $P^2=P_\mu{}P^\mu$. Of course, we will ignore spacelike momenta and the constraints of the model are therefore
$P^2\geq0$
and the skewness of $S$.
\subsection{General conservation laws}

If the equations of motion (\ref{dotPbis}) and (\ref{dotSbis}) hold, Equation (\ref{TCbis}) giving the expression of the matter distribution yields furthermore \cite{Duv72,Sou74}
\begin{equation}
\la\cT_C,L_Z\rg\ra=\int_{C}{d(\Psi(Z))}
\label{intCdPsi(Z)}
\end{equation}
for all vector field $Z$ of $M$, where
\begin{equation}
\Psi(Z)=P_\mu{}Z^\mu+\half{}S^{\mu\nu}\nabla_\mu{}Z_\nu.
\label{Psi}
\end{equation}

Therefore, if $Z$ happens to be a Killing vector field of $(M,\rg)$, i.e., if $L_Z\rg=0$, then~(\ref{intCdPsi(Z)}) readily entails that
\begin{equation}
\Psi(Z)=\const
\label{Psibis}
\end{equation}
is a \textit{first-integral} of the MPD equations (\ref{dotP}) and (\ref{dotS}). 

\goodbreak

The ``moment map'', $\Psi$, of the iso\-metries yields therefore conserved quantities for spinning particles freely falling in the gravitational field.
Let us emphasize that Equation~(\ref{Psi}) provides \textit{universal conservations laws}, independently of any choice of equation of state such as , e.g., (\ref{SP=0}). They will prove useful in the next sections where spacetime will admit a large group of isometries.

\subsection{The Souriau-Saturnini equations}\label{SouriauSaturniniSection}

The system (\ref{dotP}) and (\ref{dotS}) is non-deterministic as it clearly lacks an expression for the velocity, $\dX$. In order to cure this indeterminacy, one has to impose \textit{equations of state} (or \textit{supplementary condition}); one usually posits \cite{Tul59,Dix70,Sou74,Ste78,HLGBN}\footnote{We favor the Tulczyjew constraint (\ref{SP=0}) instead of the Pirani condition $S\dX=0$ for the following geometrical reason. It does yield the standard $6$-dimensional projective twistor spaces $\bbP\bbT^3_\pm$ as the symplectic spaces of motions for spinning massless particles with helicity (\ref{s}) in Minkowski spacetime. On the other hand, the second condition yields another, $10$-dimensional, coadjoint orbit of $\SU(2,2)$ featuring, in flat spacetime, extra degrees of freedom whose physical interpretation remains uncertain; see, e.g., \cite{DF78}.}
\begin{equation}
SP=0
\label{SP=0}
\end{equation}
which readily implies 
\begin{equation}
\Tr(S^2)=\const.
\label{P2S2const}
\end{equation}
We will furthermore show in Appendix \ref{AppendixA} that the property of $P^2$ to be also a constant of the motion, namely 
\begin{equation}
P^2=\const
\label{P2const}
\end{equation}
is compatible with the equations of motion of MPD and the equation of state  (\ref{SP=0}) provided the following consistency condition holds everywhere on the worldline, namely
 \begin{equation}
 P^2+\frac{1}{4}\Tr(S R(S))\neq0.
\label{CompatibilityCondition}
\end{equation}
We may therefore promote these constants of the motion to constants of the system, and consistently describe \textit{photons} with the two further constraints 
\begin{equation}
P^2=0
\qquad
\&
\qquad
-\half\Tr(S^2)=s^2
\label{ZeroMass+Spin}
\end{equation}
on mass and spin, where
\begin{equation}
s=\pm\hbar
\label{s}
\end{equation}
denotes the scalar spin; the helicity (handedness) of the photon is $\sign(s)$.

\goodbreak

The choice (\ref{SP=0}) finally enables us to determine the velocity and, accordingly, the full system of \textit{Souriau-Saturnini equations} of motion which takes the very particular form \cite{Sat76}
\begin{eqnarray}
\label{dotXter}
\dX&=&P+\frac{2}{R(S)(S)}S R(S) P\\[4pt]
\label{dotPter}
\dP&=&-s\,\frac{\Pf(R(S))}{R(S)(S)}\,P\\[4pt]
\label{dotSter}
\dS&=&P\overline{\dX}-\dX\barP
\end{eqnarray}
where\footnote{\label{FootnoteDeleocalization} In the sequel, we will assume $R(S)(S)\neq0$ so as to handle a nonsingular system accounting for localized test particles; see (\ref{RSSneq0}). Let us recall that, on the contrary, massless spinning particles are delocalized in (flat) Minkowski spacetime where they dwell on null affine $3$-planes, $P^\perp$, instead of worldlines \cite{SSD}. Consequently, although special relativity forbids a non-vanishing  spin component transverse to the spatial momentum (spin is enslaved), the gravitational field in general relativity cures that difficulty as long as (\ref{RSSneq0}) holds.} 
\begin{equation}
R(S)(S):=-\Tr(SR(S))=R_{\mu\nu\alpha\beta}\,S^{\mu\nu}S^{\alpha\beta}.
\label{DefRSS}
\end{equation}
In (\ref{dotPter}), the Pfaffian of a skewsymmetric linear map, e.g., $F=R(S)$ as defined in (\ref{R(S)}), is such that $(\star\,F)F=F(\star\,F)=\Pf(F)\,\bone$, where~$\star$ is the Hodge star; we furthermore have $\det(F)=-\Pf(F)^2$. 
Let us note that alternatively $\Pf(F)=-\frac{1}{8}\sqrt{-\det(\rg_{\alpha\beta})}\,\varepsilon_{\mu\nu\rho\sigma}F^{\mu\nu}F^{\rho\sigma}$ with $\varepsilon_{\mu\nu\rho\sigma}$ the Levi-Civita symbol such that $\varepsilon_{1234}=1$. 
A proof of (\ref{P2S2const}) and of the above equations of motion (\ref{dotXter}) and (\ref{dotPter}), specific of the massless case, is given in Appendix~\ref{AppendixA}.

We note for further use that Equations (\ref{SP=0}), (\ref{ZeroMass+Spin}) and (\ref{dotXter}) imply that the momentum $P$ is orthogonal to the velocity $\dX$, i.e.,
\begin{equation}
\barP\dX=0.
\label{PperpdotX}
\end{equation}

The above equations of motion have actually another equivalent geometrical origin: they are given by the characteristic distribution of a presymplectic $2$-form on the $9$-dimensional ``evolution space'' consisting of all triples $(X,P,S)$ satisfying the following constraints: the skewsymmetry of the spin operator, the orthogonality of spin and momentum (\ref{SP=0}) and the masslessness of the particle of fixed spin (\ref{ZeroMass+Spin}). Here, we will not need to rely on these results. See, however, \cite{Kun72,DFS72,Sou74,Sat76} for a complete account of this model.

\goodbreak

\section{Photons in Robertson-Walker backgrounds}\label{PhotonsRWSection}


The metric is given in a spatial Cartesian coordinate patch $(x^1=x,x^2=y,x^3=z)$ by
\begin{equation}
\rg = -a(t)^2\left[\frac{dx^2+dy^2+dz^2}{b(x,y,z)^2}\right]+dt^2
\qquad
\&
\qquad
b(x,y,z)=1+\frac{K}{4}(x^2+y^2+z^2)
\label{gbis}
\end{equation}
where $a$ is a smooth strictly positive function of the time axis parametrized by~$x^4=t$ (cosmic time), and $K\in\bbR$ is related to the spatial scalar curvature $R^{(3)}=6K$. We will use the coordinate patch where $b$ does not vanish.

\goodbreak

The non-zero Christoffel symbols read
\begin{equation}
{\Gamma^i}_{ii}=-\frac{K x^i}{2b},
\quad
\displaystyle
{\Gamma^i}_{jj}=\frac{K x^i}{2b},
\quad
\displaystyle
{\Gamma^i}_{ij}=-\frac{K x^j}{2b},
\quad
{\Gamma^i}_{i4}=\frac{a'}{a},
\quad
{\Gamma^4}_{ii}=\frac{aa'}{b^2}
\label{GammaBis}
\end{equation}
for all $i,j=1,2,3$ and $i\neq{}j$. 

The Ricci tensor reads 
$\Ric=(2 K + 2a'^2 +aa'')/b^2\left(dx^2+dy^2+dz^2\right)-3(a''/a)dt^2$,
and the scalar curvature is
$R=-6(K + a'^2 +aa'')/a^2$. Spacetime $(M,\rg)$ is conformally flat.

\subsection{The general equations of motion of photons}
\label{U6}

In the above coordinate system, the (future pointing) null linear momentum of the particle is written as
\begin{equation}
P=\left(
\begin{array}{c}
\displaystyle
\frac{\hbp}{a}\\[8pt]
\displaystyle
 \frac{\nhp}{b}
\end{array}\right)
\label{Pbis}
\end{equation}
with $\hbp\in\bbR^3\setminus\{0\}$, the spatial linear momentum, and $\nhp\dpp=\sqrt{\hbp\cdot\hbp}$ (positive energy).\footnote{The dot and cross products are those of Euclidean space $\bbR^3$, the superscript ``$T$'' in (\ref{Sbis}) means transposition, and we have the linear map $j(\bs):\hbp\mapsto\bs\times\hbp$.}

\goodbreak

Accordingly, the spin tensor reads
\begin{equation}
S=\left(
\begin{array}{cc}
j(\bs)&\displaystyle
-\frac{(\bs\times\hbp)}{\nhp}\frac{b}{a}\\[6pt]
\displaystyle
-\frac{(\bs\times\hbp)^T}{\nhp}\frac{a}{b}&0
\end{array}\right)
\label{Sbis}
\end{equation}
where the spin 
vector $\bs\in\bbR^3\setminus\{0\}$ satisfies, along with $P$, the constraints (\ref{SP=0}) and (\ref{ZeroMass+Spin}); the latter yields the scalar spin 
\begin{equation}
s=\frac{\bs\cdot\hbp}{\nhp}
\label{sbis}
\end{equation}
(not to be confused with the norm $\Vert\bs\Vert$ of the spin vector) satisfies (\ref{s}) for photons.

\goodbreak

Some more calculation yields
\begin{equation}
R(S)=-\frac{2}{a^2}\left(
\begin{array}{cc}
\displaystyle
(K+a'^2)j(\bs)&
\displaystyle
-\frac{\bs\times\hbp}{\nhp}\,b\,a''\\[6pt]
\displaystyle
-\frac{(\bs\times\hbp)^T}{\nhp}\,\frac{a^2a''}{b}&0
\end{array}\right)
\label{RSbis}
\end{equation}
together with $\det(R(S))=0$, hence
\begin{equation}
\Pf(R(S))=0
\label{PfRS}
\end{equation}
which will entail, via Equation (\ref{dotPter}), that the linear momentum, $P$, is actually parallel-transported.

We furthermore find
\begin{equation}
R(S)(S)=\frac{4}{a^2}\left(\Vert\bs\Vert^2(aa''-(K+a'^2))-s^2\,aa''\right).
\label{RSSbis}
\end{equation}
Also do we get
\begin{equation}
S R(S)P=\frac{2}{a^2}((K+a'^2)-aa'')\left(\Vert\bs\Vert^2 P-s\,\hW\right)
\label{SRSPbis}
\end{equation}
where
\begin{equation}
\hW=\nhp\left(
\begin{array}{c}
\displaystyle
\frac{\bs}{a}
\\[10pt]
\displaystyle
\frac{s}{b}
\end{array}\right)
\label{Wbis}
\end{equation}
is interpreted as the \textit{polarization vector} of the massless particle in this gravitational field. 


With these preparations, extra work is needed to express the original equation (\ref{dotXter}) for the velocity in terms of that involving the natural parameter: cosmic time, $t$. We find
\begin{equation}
\frac{dX}{\!d\tau}=-\frac{4s^2\nhp}{R(S)(S)}\frac{(K+a'^2)}{a^2b}\frac{dX}{\!dt}
\label{dX/dtau=dt/dtau.dX/dt}
\end{equation}
with 
\begin{equation}
\frac{dX}{\!dt}=\frac{aa''b}{\nhp(K+a'^2)}\left[P-\left(1-\frac{K+a'^2}{aa''}\right)\frac{\hW}{s}\right]
\label{dX/dt}
\end{equation}
where the vector $\hW$, given by Equation (\ref{Wbis}), features the polarization-driven \textit{anomalous velocity}. Let us anticipate that, by the Friedman equations, $K+a'^2\neq0$.

We however mention that the reparametrization $\tau\mapsto{}t$ in Equation (\ref{dX/dtau=dt/dtau.dX/dt}) will be considered legitimate provided it be a diffeomorphism, $d\tau/dt\neq0$, i.e.,
\begin{equation}
R(S)(S)\neq0
\label{RSSneq0}
\end{equation}
which is assumed to hold for the time being. We will check this inequality in Section \ref{NumericalSolutionsSection} by numerically integrating the equations of motion (\ref{dx})--(\ref{ds}) in the particular case of the flat Lambda Cold Dark Matter model.


\goodbreak

We note that $\barP{}\hW=0$ and 
$\hW^2
=
-\nhp^2\Vert\bs^\perp\Vert^2/b^2\leq0
$
where the transverse spin $\bs^\perp$ has been defined by 
\begin{equation}
\bs=s \frac{\hbp}{\nhp}+\bs^\perp
\label{sperp}
\end{equation}
in accordance with (\ref{sbis}).
This implies that the spinning photon travels at a speed greater than the speed of spinless light (\textit{tachyonic velocity}) since
\begin{equation}
\left(\frac{dX}{\!dt}\right)^2
=
-\left[\frac{aa''-(K+a'^2)}{K+a'^2}\right]^2\,\frac{\Vert\bs^\perp\Vert^2}{s^2}.
\label{tachyon}
\end{equation}
Notice that the RHS of Equation (\ref{tachyon}) vanishes when the spin is ``enslaved'', $\bs^\perp=0$, i.e., when~$\hW$ becomes the Pauli-Lubanski vector \cite{SSD,EDHZ16}.


\goodbreak

\subsection{``Energy'' of polarized light rays}

Moreover, the vector field 
$
\Theta=a(t)\partial/\partial t
$
is clearly a conformal-Killing vector field;
it is the ``cotemperature'' vector field considered in \cite{Sou75}.
One checks that 
\begin{equation}
\nabla_\mu\Theta_\nu=a'(t)\,\rg_{\mu\nu}
\label{nablaTheta}
\end{equation}
implying that $\Theta$ is curlfree and geodesic. It is moreover timelike since $\Theta^2=a(t)^2>0$, and clearly future pointing (the time arrow).

The ``energy'' of the photon relatively to $\Theta$ is readily defined by
\begin{equation}
\cE=\barP\Theta.
\label{E}
\end{equation}
Let us show that it is actually a constant of the motion,
\begin{equation}
\cE=\const.
\label{E=const}
\end{equation}
We indeed have $\dot{\cE}=P_\mu\dot{\Theta}^\mu$ since $\dP=0$, and thus $\dot{\cE}=P^\mu\dX^\nu\nabla_\nu\Theta_\mu=a'(t)P_\mu\dX^\mu$ in view of (\ref{nablaTheta}). At last, Equation (\ref{PperpdotX}) implies $\dot{\cE}=0$.\footnote{Equation (\ref{E}) may be viewed, thanks to (\ref{nablaTheta}), as a special case of the moment map (\ref{Psi}) adapted to the conformal-Killing vector field $Z=\Theta$.} We will therefore call $\cE$ the conserved ``energy''.

\subsection{Photons in flat Robertson-Walker backgrounds}
 
From now on, we put $K=0$ and express the system (\ref{dotXter})--(\ref{dotSter}) of  equations of motion for our massless spinning particle (cf. Section \ref{SouriauSaturniniSection})  in the natural coordinate system
\begin{equation}
X=\left(
\begin{array}{c}
\bx\\
t
\end{array}\right)
\label{X}
\end{equation}
with  the comoving Euclidean coordinates $\bx\in\bbR^3$ and the cosmic time $t$.





\goodbreak

$\bullet$
Equation (\ref{dX/dt}) for the velocity with respect to cosmic time readily yields  the expression of the spatial velocity 
\begin{equation}
\frac{d\bx}{dt}
=
\frac{a''}{a'^2}\frac{\hbp}{\nhp}+\frac{1}{a}\left[1-\frac{aa''}{a'^2}\right]\frac{\bs}{s}.
\label{dx}
\end{equation}


$\bullet$ Likewise, we find 
that Equations (\ref{dotPter}) and (\ref{PfRS}) yield, together with the expression (\ref{GammaBis}) of the Christoffel symbols, the equation governing the time evolution of the spatial momentum in (\ref{Pbis}), namely 
\begin{equation}
\frac{d\hbp}{dt}
=
-\frac{a'}{a}\left[\frac{aa''}{a'^2}\,\hbp+\nhp\left(1-\frac{aa''}{a'^2}\right)\,\frac{\bs}{s}\right].
\label{dp}
\end{equation}
Note that the equation for the evolution of $P^4=\Vert\hbp\Vert$, resulting from (\ref{dotPter}), is clearly compatible with Equation (\ref{dp}).


$\bullet$ Equation (\ref{dotSter}) for the spin evolution can now be read off, with the help of (\ref{Sbis}) and (\ref{dx}), as 
\begin{equation}
\frac{d\bs}{dt}
=
-\left(1-\frac{aa''}{a'^2}\right)\frac{\bs}{s}\times\hbp
-\frac{a'}{a}\,\bs
+\frac{a'}{a}\left[\frac{\ns^2}{s}\left(1-\frac{aa''}{a'^2}\right)+s\,\frac{aa''}{a'^2}\right]\frac{\hbp}{\nhp}.
\label{ds}
\end{equation}


%

The non-linear coupled system (\ref{dx})--(\ref{ds}) constitutes precisely the deterministic set of ordinary differential equations that we will be studying from now on to describe massless particles with spin $s$ in a (spatially) flat Robertson-Walker background.

\goodbreak

\subsection{Conservation laws}


The Euclidean group $\rE(3)=\rO(3)\ltimes\bbR^3$ being a group of isometries, its generators are Killing vector fields of the metric (\ref{gbis}), viz.,
$Z=\left(\varepsilon^i_{\,jk}\,\omega^jx^k+\gamma^i\right)\,\partial/\partial x^i$,
where $\bomega,\bgamma\in\bbR^3$ stand for infinitesimal rotations and translations, respectively; the $\varepsilon^i_{\,jk}$ are the structure constants of $\so(3)$.

Using the general expression (\ref{Psi}) of the moment map, $\Psi$, associated with a Killing vector field, $Z$, together with the expressions (\ref{Pbis}) and (\ref{Sbis}) for $P$ and $S$, we find in a straightforward fashion
$
\Psi(Z)=-\bcL\cdot\bomega-\bcP\cdot\bgamma
$
where 
\begin{equation}
\bcP=a(t)\,\hbp+\frac{a'(t)}{\nhp}\,\bs\times\hbp
\label{cP}
\end{equation}
stands for the conserved linear momentum and
\begin{equation}
\bcL=\bx\times\bcP+\bs
\label{cL}
\end{equation}
for the conserved angular momentum featuring an extra spin contribution.

\goodbreak


In addition, the constant ``energy'' (\ref{E}) associated with the cotemperature vector field of the system then reads
\begin{equation}
\cE=a(t)\,\nhp=\const
\label{Energy}
\end{equation}
and verifies $\cE>0$ since we have chosen $P$ to be future pointing.

\goodbreak


Moreover we find that $\Vert\bcP\Vert^2-\cE^2=a'^2\Vert\bs\times\hbp\Vert^2/\nhp^2=a'^2\Vert\bs^\perp\Vert^2$, and therefore
\begin{equation}
a'(t)\Vert\bs^\perp\Vert=\const
\label{apsperp=const}
\end{equation}
is a new first integral of our system associated with transverse spin (\ref{sperp}).


\section{Numerical solutions}\label{NumericalSolutionsSection}

In this section we numerically solve the equations of motion for a massless particle with non-vanishing spin in flat Robertson-Walker spacetime with a given scale factor $a(t)$, that we suppose increasing with cosmic time $t$. To be specific, we take the Friedman solution of the flat Lambda Cold Dark Matter model, namely
\begin{align}
a(t)\,=\,a_0\lp\frac{\cosh[\sqrt{3\Lambda }\,t]-1}{\cosh[\sqrt{3\Lambda}\,t_0]-1}\rp^{1/3}
\label{scalef}
\end{align}
where $\Lambda$ stands for the cosmological constant. We have chosen the origin of cosmic time at the big bang, $t=0$, and note $a_0\dpp=a(t_0)$ the scale factor today. This model is the standard model of cosmology and its successes and draw-backs are well reviewed in \cite{pdg}.

\goodbreak

Our task is to solve the system (\ref{dx}), (\ref{dp}), (\ref{ds})
of nine first order, ordinary dif\-ferential equations for nine unknowns $\bx(t)$, $\hbp(t)$, $\bs(t)$. We choose the nine initial conditions at $t=t_e$, the time of emission (or creation) of the massless particle.
We orient our Cartesian coordinates such that
\bb \bx_e=0,\qq
\hbp_e=\begin{pmatrix}
\nhpe\\0\\0
\end{pmatrix},\qq
\bs_e=\begin{pmatrix}
s\\s^\perp_e\\0
\end{pmatrix}
\label{initial}
\ee
with $s_e^\perp\dpp=\Vert\bs_e^\perp\Vert\geq0$.

Of course, we will use the six conserved quantities  $\bcL$ and $\bcP$, Equations (\ref{cL}) and~(\ref{cP}), to eliminate the unknowns $\bs(t)$ by
\bb 
\bs(t)=\begin{pmatrix}
s&\!\!\!\!+a'_e\,s^\perp_e\,x^2(t)&\\[2mm]
s^\perp_e&\!\!\!\!-a'_e\,s^\perp_e\,x^1(t)&\!\!\!{-}\cE\,x^3(t)\\[2mm]
&&\!\!\!{+}\cE\,x^2(t)
\end{pmatrix}
\label{bst}
\ee
and $\hbp(t)$ by
\begin{align}
\hbp(t)=&\ \frac{\cE\,a(t)^{-1}}{1+(a'(t)^2\,s^2+a_e^{'2}\,s^{\perp\,2}_e)/\cE^2}\,\nonumber\\[2mm]&\cdot \begin{pmatrix}
1&\!\!\!\!+a'_e\,a'(t)\,s^\perp_e\,s_2(t)/\cE^2 &\!\!\!\!+a'(t)^2s\,s_1(t)/\cE^2\\[2mm]
{-}a'(t)\,s_3(t)/\cE&\!\!\!\!-a'_e\,a'(t)\,s^\perp_e\,s_1(t)/\cE^2 &\!\!\!\!+a'(t)^2s\,s_2(t)/\cE^2\\[2mm]
\ {}a'(t)\,s_2(t)/\cE&\!\!\!\!{-}a'_e\,s_{2e}/\cE &\!\!\!\!+a'(t)^2s\,s_3(t)/\cE^2
\end{pmatrix}.
\label{bpt}
\end{align}
To obtain the last equation we have used the matrix inverse
\bb
[\bone+j(\mathbf{z})]^{-1}\,=\,\frac{1}{1+\Vert\mathbf{z}\Vert^2}\, 
[\bone-j(\mathbf{z}) +\mathbf{z}\,\mathbf{z}^T]
\ee
for all $\bz\in\bbR^3$, and the ``conservation'' law (\ref{apsperp=const}) of the transverse component of the spin:
\bb
 a'(t)\,s^\perp(t) = a'_e \,s^\perp_e
\ee
 with $s^\perp(t)\dpp=\sqrt{\Vert\mathbf{s}(t)\Vert^2-s^2}$, and $a'_e\dpp=a'(t_e)$.
 
\goodbreak
 
The remaining equation of motion (\ref{dx}) reads:
\bb 
\bx'(t)=\,\frac{a(t)\, a''(t)}{a'(t)^2}\,\frac{\hbp(t)}{\cE}\, +
\lp\frac{1}{a(t)}\, -\,\frac{a''(t)}{a'(t)^2}\rp\,\frac{\bs(t)}{s}
\label{three}
\ee
where $\bp(t)$ and $\bs(t)$ are as in (\ref{bpt}) and (\ref{bst}), respectively, $\cE$ being given by (\ref{Energy}).

We integrate them using a Runge-Kutta algorithm. To specify our initial conditions (\ref{initial}), we must choose units. We have already put $c=1$ and we add $\hbar=1$ and the Hubble constant $H_0=1$. These three choices determine the units of time, length and mass, which we call astro-seconds [as], astro-meter [am] and astro-gram [ag]:
\begin{align}
{\rm as} &= 13.97/h_{70}\,{\rm Gyrs} = 4.408/h_{70}\cdot10^{17}\,{\rm s},\\
{\rm am} & = 1.321/h_{70}\cdot10^{26}\,{\rm m},\\
{\rm ag} & =2.664\,h_{70}\cdot10^{-69}\,{\rm kg},
\end{align}
with $h_{70}\dpp=h/0.70$ and $h=0.673\pm0.012$ today \cite{pdg}.
In these units we have $H_0=1/{\rm as}$, $c=1\,{\rm am}/{\rm as}$, $\hbar=1\, {\rm ag}\,{\rm am}^2/{\rm as}$, the cosmological constant is $\Lambda = 3\cdot(0.685\pm0.017)\, {\rm am}^{-2}$ and the age of the universe $t_0=0.951\,{\rm as}$. (For completeness we record Newton's constant, $G=1.49868\cdot h_{70}^2\cdot 10^{-122}\,{\rm am}^3\,{\rm as}^{-2}\,{\rm ag^{-1}}$, which we do not use explicitly.) Quantum mechanics tells us that spin one particles have $\Vert\bs_e\Vert^2=2 \hbar^2$ and $s_{1e}=s=0,\pm\,\hbar$. For massless particles the projection of the spin onto the momentum direction, i.e., the scalar spin, is maximal or minimal, $s=\pm\,\hbar$ and therefore 
 $s^\perp_e=\hbar$. (A graviton would have $s=\pm\,2\, \hbar$ and $s^\perp_e=\sqrt{2}\,\hbar$.)
 
\begin{figure}[h]
\begin{center}
\includegraphics[scale=0.8]{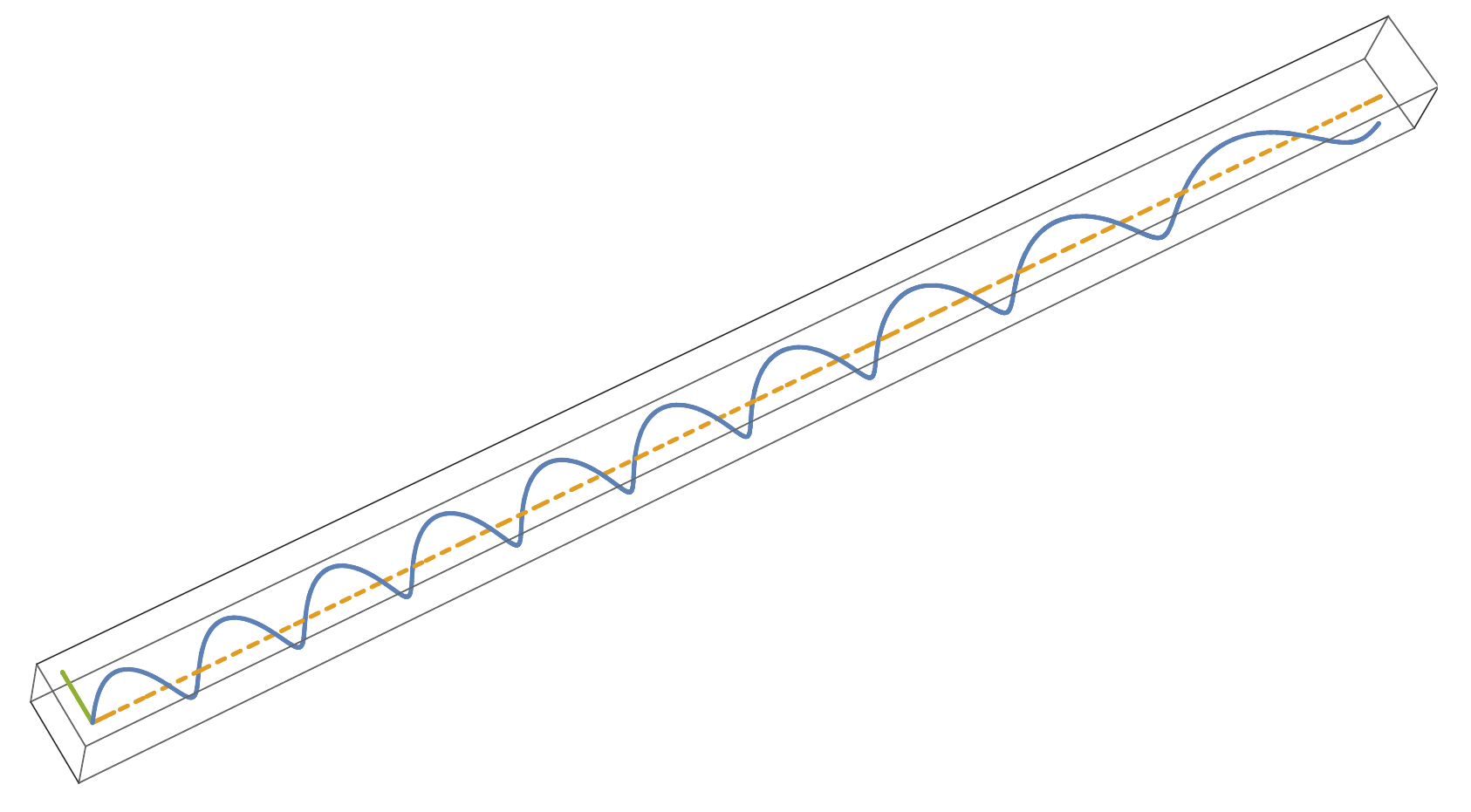}
\caption {The trajectory of the the photon, $\bx(t)$, in comoving coordinates is the helix. The dashed line is the geodesic followed by mass- and spin-less test particles. The initial transverse direction of the spin, i.e. the $x^2$-direction is indicated by the short line at the left.\label{helix}}
\end{center}
\end{figure}

For concreteness, consider a photon from the Lyman $\alpha$ emission line, (period  $T = 4\cdot 10^{-16}\, {\rm s}  = 8.72 \cdot 10^{-34} \,{\rm as}$, wavelength $\lambda =1.2\cdot 10^{-7}\,{\rm m}= 8.72 \cdot 10^{-34} \,{\rm am}$)  and with redshift $z=2.4$. From $z+1=a_0/a(t_e)$ and by numerically inverting the scale factor $a(t)$, Equation (\ref{scalef}), we obtain its emission time
$t_e=0.188\,{\rm as}$ and then $a'_e=1.06$.  The critical time, $a''(t_\mathrm{crit})=0$, is $t_\mathrm{crit}=0.530\,{\rm as}$. The photon energy at emission is $\nhpe=2\pi\hbar /T$ and the conserved ``energy'' $\cE=a_e\,\nhpe=a_0\,2\pi\hbar /T/(z+1)$.

\goodbreak


Of course, a numerical solution with the initial condition for $p_1$ as large as $10^{34}$ does not work and we will start with a modest period of $T=1.2\cdot10^{-2}\,{\rm as}$ 
instead of that of Lyman $\alpha$. 

For consistency we must first check that $R(S)(S)$ in the denominator of the equations of motion (\ref{dotXter}) and (\ref{dotPter}) does not vanish in the domain of numerical integration. We use the form (\ref{RSSbis}) with $K=0$ for $R(S)(S)$ and find that it remains positive for the numerical solution of the trajectory $\bx(t)$, justifying the assumption (\ref{RSSneq0}).

Figure \ref{helix} shows the trajectory of the photon $\bx(t)$ in comoving coordinates. It is a helix around the geodesics of  mass- and spin-less test particles  or equivalently around the trajectory of `photons' with alined spin, $\bs^\perp  =0$. The periods of the cycles of the helix vary but remain of the order of the period $T$ of the photon at emission.

\begin{figure}[h]
\begin{center}
\includegraphics{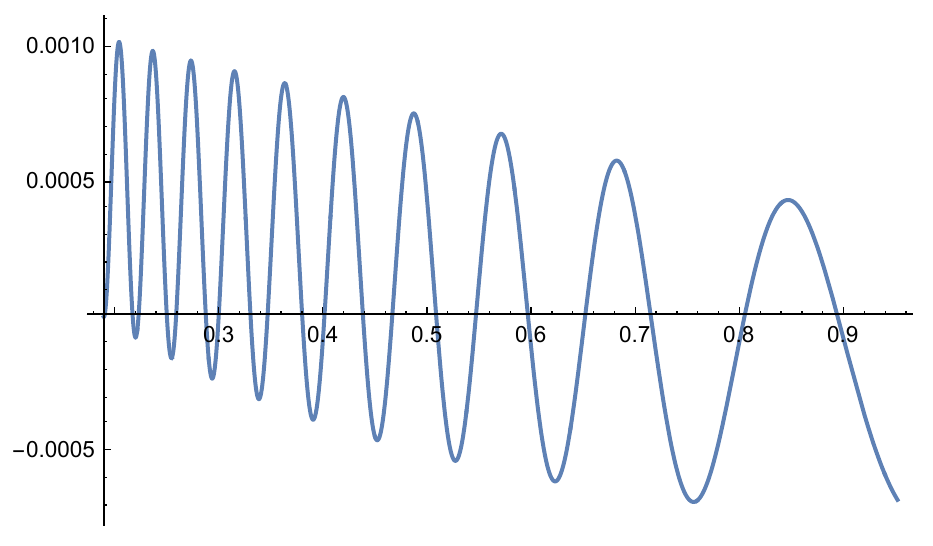}
\caption 
{Details of the race between our photon and a mass- and spin-less test particle: $x^1(t)-x^1(t)|_{x^\perp=0}$ as a function of cosmic time. 
\label{xL}}
\end{center}
\end{figure}

The particular form of the helix guarantees that, although being a tachyonic curve, it does propagate with the speed of light as can be seen from Figure \ref{xL}. While both coordinate functions, $x^1(t)$ of the photon and $x^1(t)|_{x^\perp=0}$ of the light-like geodesic, vary monotonically from 0 to 1.32, their differences, shown in Figure \ref{xL}, remain small. 

The transverse (comoving) coordinate of the helix $x^\perp(t)\dpp=\sqrt{x^2(t)^2+x^3(t)^2}$ is shown in Figure \ref{xT}. It varies, but remains of the order of the wavelength of the photon at emission.  These variations vanish around  $t_{\rm vanish}=0.66\, {\rm as}$, corresponding to a redshift $z_{\rm vanish}=1.6$. We will see in the next section that  $t_{\rm vanish}$ is determined by $a'_e\,x^1(t_{\rm vanish})=1$.

\begin{figure}[hp]
\begin{center}
\hspace{2.2mm}
\includegraphics[scale=1]{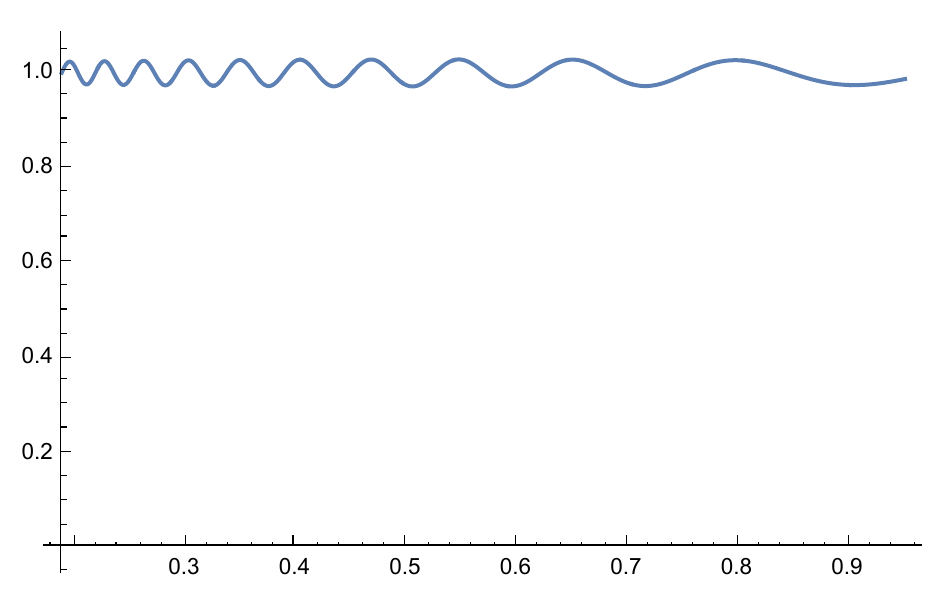}
\includegraphics[scale=1]{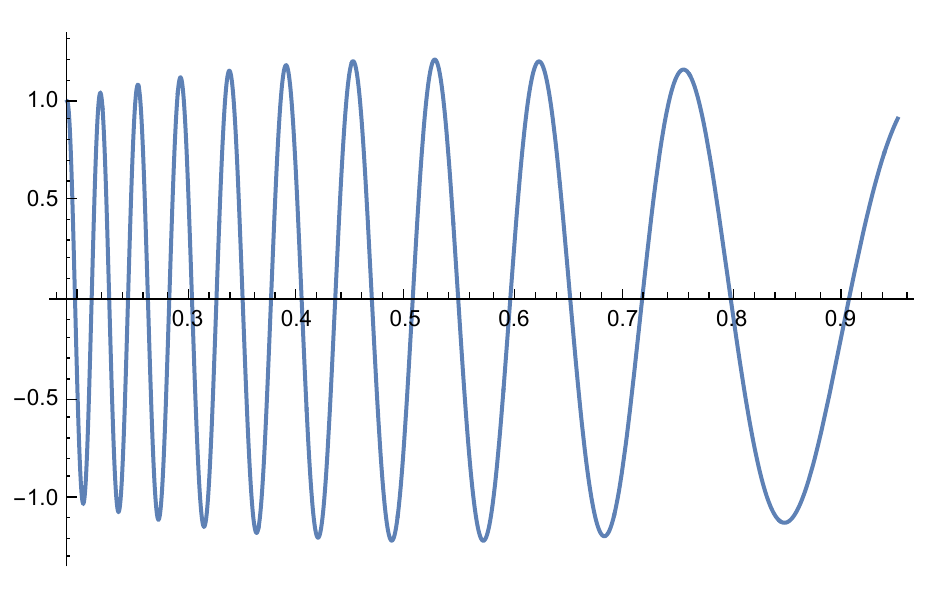}
${}$\hspace{2mm}
\includegraphics[scale=1]{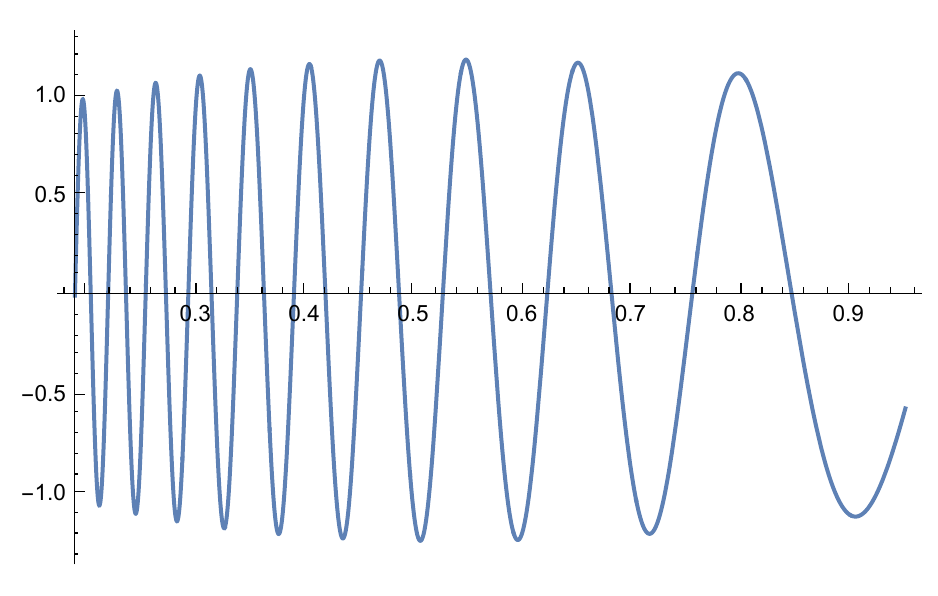}
\caption 
{The three spin components $s_1(t)$, $s_2(t)$, and $s_3(t)$ as a function of cosmic time.  \label{spin}}
\end{center}
\end{figure}

The spin vector $\bs(t)$ rotates in step with the cycles of the helix, see Figure \ref{spin}. Its rotation axis deviates only slightly from the $x^1$-axis.

Naturally, we want to know the evolution of the variable periods of the cycles of the helix and compute  its center and radius in the limit of a small atomic period $T$ or equivalently of small spin. We do this in the next section by a perturbative computation, which will also tell us that our numerical result is generic for small spins. The extension to large spin motivated by macroscopic electromagnetic wavepackets is however not included.


\section{Perturbative solutions}\label{PerturbativeSolutionsSection}

In this section we return to flat Robertson-Walker cosmologies. The scale factor $a(t)$ is arbitrary up to the following constraints that are to be satisfied for all cosmic times $t$ between emission and today, $t_e\le t\le t_0$:
\bb
a>0, \qq a'>0, \qq s^{\perp 2}\,a\,a''-\Vert\mathbf{s}\Vert^2 a^{'2}\,\not=0, \qq 1/a-a''/a^{'2}\not=0.
\label{constraints}
\ee

Let us write out our three equations of motion (\ref{three}) for $\bx(t)$ by explicitly eliminating $\bs(t)$ and $\bp(t)$. We are dealing with two small parameters of the order of $10^{-34}$ from the longitudinal and transverse initial values of the spin:
\bb \eta\dpp =\,\frac{s}{\cE}\, =\,\frac{\pm T}{2\pi \,a_e}\, ,\qq
\epsilon\dpp =\,\frac{s^\perp_e}{\cE}
\label{etaepsilon}
\ee
as given by (\ref{initial}). Let us recall that $T$ is the atomic period at emission. We also choose the normalization $a_0=1\,{\rm am}$.

We consider $\eta$ to be a fixed, non-zero number and $\epsilon$ to vary between 0 and $|\eta|$. Indeed we know that for $\epsilon =0$ our trajectory is the light-like geodesic and we want to know how the trajectory of the photon deviates from this geodesic to first order in $\epsilon$. At the end of our calculation we will put $\epsilon =| \eta|$ for the photon. To alleviate notations we will often suppress the argument~$t$ and, as before, denote evaluation at initial time $t_e$ simply by the subscript ``$e$''. From Equations (\ref{three}), (\ref{bpt}) and (\ref{bst}) we obtain:
\begin{align}
\frac{dx^1}{\!\!dt}
=&\,\frac{a''}{a^{'2}\,(1+a_e^{'2}\,\epsilon^2+a^{'2}\,\eta^2)}\,\nonumber \\[2mm]
&\cdot\lb 1+a'_ea'\,\epsilon^2-a^{'2}_ea'x^1\,\epsilon^2{-}a'_ea'x^3\,\epsilon+a^{'2}\,\eta^2+a'_ea^{'2}x^2\,\eta\epsilon\rb\nonumber\\[2mm]
&\qq+\lp\,\frac{1}{a}\, -\,\frac{a''}{a^{'2}}\, \rp\lb1+a'_ex^2\,\frac{\epsilon}{\eta}\,\rb\,, \\[2mm]
\frac{dx^2}{\!\!dt}
=&\,\frac{a''}{a^{'2}\,(1+a_e^{'2}\,\epsilon^2+a^{'2}\,\eta^2)}\,\nonumber \\[2mm]
&\cdot\lb -a'x^2-a'_ea'\,\eta\epsilon-a^{'2}_ea'x^2\,\epsilon^2
+a^{'2}\,\eta\epsilon-a'_ea^{'2}x^1\,\eta\epsilon{-}a^{'2}x^3\,\eta
\rb\nonumber\\[2mm]
&\qq+\lp\,\frac{1}{a}\, -\,\frac{a''}{a^{'2}}\, \rp\lb\,\frac{\epsilon}{\eta}\, -a'_ex^1\,\frac{\epsilon}{\eta}\,{-}x^3\,\frac{1}{\eta}\, \rb\,, \\[2mm]
\frac{dx^3}{\!\!dt}
=&\,\frac{a''}{a^{'2}\,(1+a_e^{'2}\,\epsilon^2+a^{'2}\,\eta^2)}\,\nonumber \\[2mm]
&\cdot\lb{-} a'_e\,\epsilon {+}a'\,\epsilon{-}a'_ea'x^1\,\epsilon-a'x^3{+}a^{'2}x^2\,\eta
\rb\nonumber\\[2mm]
&\qq{+}\lp\,\frac{1}{a}\, -\,\frac{a''}{a^{'2}}\, \rp x^2\,\frac{1}{\eta}\, . 
\end{align} 
Note that here, thanks to the cosmological principle and the symmetries of the model, the equations of motion are affine.

In the limit $\epsilon\rightarrow 0$ the two transverse coordinates $x^2$ and $x^3$ vanish identically and we will suppose that the two limits of $x^2/\epsilon$ and of $x^3/\epsilon$ exist and call them $y_2(t)$ and $y_3(t)$. This hypothesis will be supported by the comparison of our perturbative results with the numerical ones of the preceding section.

\begin{figure}[h]
\begin{center}
\includegraphics{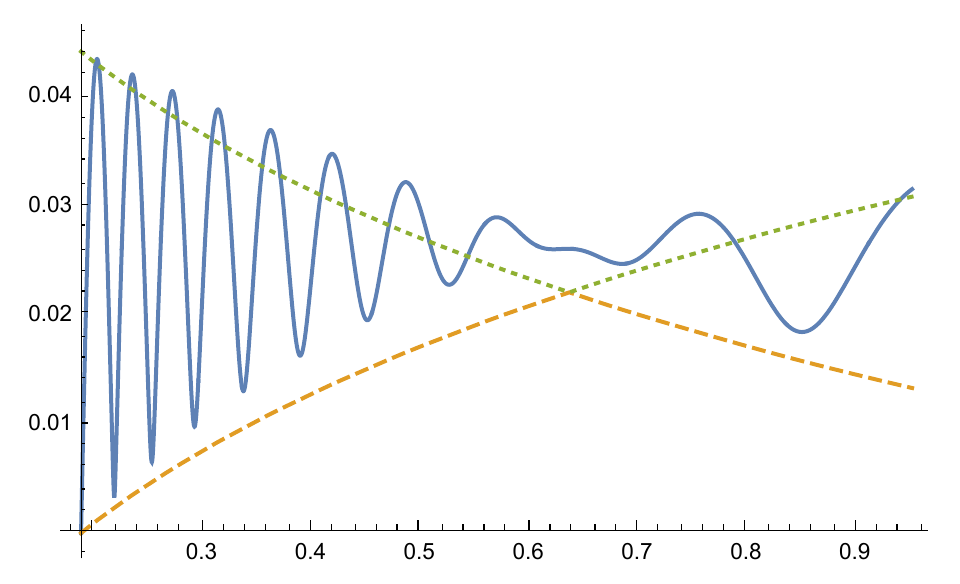}
\caption 
{The transverse coordinate of the helix $x^\perp(t)\dpp=\sqrt{x^2(t)^2+x^3(t)^2}$ as a function of cosmic time. The perturbative bounds of Equation (\ref{bounds}) are indicated by  dashed and dotted lines. They touch at $t_{\rm vanish}$. \label{xT}}
\end{center}
\end{figure}

Then, to leading order in $\eta$ and $\epsilon$, the equations of motion read:
\begin{align}
\frac{dx^1}{\!\!dt}
\sim&\,\frac{1}{a}\, +a'_e\lp\,\frac{1}{a}\, -\,\frac{a''}{a^{'2}}\, \rp\,y_2\,\frac{\epsilon^2}{\eta}\, ,\label{xp1}\\[2mm]
\frac{dy_2}{\!\!dt}
\sim&\lp\,\frac{1}{a}\, -\,\frac{a''}{a^{'2}}\, \rp\lb1-a'_ex^1{-}y_3\rb\,\frac{1}{\eta}\,, \\[2mm]
\frac{dy_3}{\!\!dt}
\sim&{}\lp\,\frac{1}{a}\, -\,\frac{a''}{a^{'2}}\, \rp\,y_2\,\frac{1}{\eta}\,.\label{xp3}
\end{align}
Equation (\ref{xp1}) may tempt us to think that $x^1(t)-x^1(t)|_{x^\perp=0}$ is of first order in $\epsilon=|\eta|$. In fact the difference is of second order: combining Equations (\ref{xp1}) and (\ref{xp3}) yields $x^1(t)-x^1(t)|_{x^\perp=0}\sim{}a'_e\,y_3(t)\,\epsilon^2$.
This is important for the race between photons and mass- and spin-less test particles; compare with Figure \ref{xL}.

Let us suppose that the scale factor is such that $1/a-a''/a^{'2}$ never vanishes (see (\ref{constraints})), which is true for the scale factor (\ref{scalef}) of the flat Lambda Cold Dark Matter model. Then, for non vanishing spin, $\eta\neq0$, we may define a new parameter $\theta  $ by 
\bb \,\frac{\de \theta }{\de t}\, =\,\frac{1}{|\eta|}\, \lp\,\frac{1}{a}\, -\,\frac{a''}{a^{'2}}\, \rp \qq {\rm and} \qq \theta (t_e)=0.\ee
To leading order, we have
\bb \theta (t)\sim\,\frac{1}{|\eta|}\, \lb x^1(t)+\,\frac{1}{a'(t)}\, -\,\frac{1}{a'_e}\, \rb\,, \ee 
and $\theta _0\dpp=\theta (t_0)$ is of the order of $10^{34}$. We also define new coordinates by $z_1(\theta )\dpp=x^1(t(\theta ))$ and $z_{2/3}(\theta )\dpp=y_{2/3}(t(\theta )) $; by abuse of notation we write $t(\theta )$ for the inverse function of $\theta (t)$. In the new coordinates the transverse equations of motion read to leading order:
\bb
\,\frac{\de z_2}{\!\!\de\theta }\, =\,{\rm sign}(\eta)\,({-}z_3+1-a'_e\, z_1),
\qq\,
\frac{\de z_3}{\!\!\de\theta }\, =\,{}{\rm sign}(\eta)\,z_2.
\ee
Let us combine these two equations:
\bb \,\frac{\de^2z_2}{\!\!\de\theta ^2}\,+z_2=- \,{a'_e}\,\frac{\eta}{1-aa''/a^{'2}}\,.\ee
To proceed, we write $z_2(\theta)=\dpp z_{20}(\theta)+\eta\,z_{21}(\theta)$ and obtain two differential equations:
\begin{align}
\,\frac{\de^2z_{20}}{\!\!\!\de\theta ^2}\,+z_{20}&=0\,,\\[2mm]
\,\frac{\de^2z_{21}}{\!\!\!\de\theta ^2}\,+z_{21}&=- \,{a'_e}\,\frac{1}{1-aa''/a^{'2}}\,.
\end{align}
The solution $z_{21}$ of the second equation does not concern us because we want $x^2=\epsilon z_2=\epsilon\,(z_{20}+\eta\,z_{21})$ only to first order.
Solving the first, homogeneous, equation with our initial condition $z_2(0)=0$, coming from the first equation in (\ref{initial}), we obtain $z_2\sim k\sin\theta $ with one integration constant $k$. Now we set $\epsilon=\vert\eta\vert$. The initial condition $z_3(0)=0$ fixes $k={\rm sign}(\eta)
$ and gives $z_3\sim{-}\cos\theta -1+a'_e\,z_1$. 

Let us denote by $T_{\rm helix}(t)$ the period -- measured in cosmic time $t$ -- of one cycle of the helix at time $t$. Measured in the parameter $\theta $ this period is $2\pi $ and independent of $\theta $. Therefore
\bb T_{\rm helix}(t)\sim2\pi \,\frac{\de t}{\de\theta }\,=\,\frac{a(t)}{a_e}\,\frac{a'(t)^2}{a'(t)^2-a(t)\,a''(t)}\,\, T=\,\frac{a(t)}{a_e}\,\frac{1}{1+q(t)}\,\, T\,,\ee
with the definition (\ref{etaepsilon}) of $\eta$, and with the expression  $q(t)\dpp =-a(t)\,a''(t)\,a'(t)^{-2}$ of the de\-celeration parameter.

At cosmic time $t$, the center of the helix has comoving coordinates 
\bb
\begin{pmatrix}
 x^1(t)\\ 0 \\ {}|\eta|\,[1-a'_e\,x^1(t)]
 \end{pmatrix}
.\ee
Its offset from the light-like geodesic vanishes at time $t_{\rm vanish}$ where the center of the helix coincides with the trajectory of the null geodesic.

The comoving radius of the helix with respect to its center is $|\eta|$ and time-independent. The time-dependence of the comoving transverse coordinate of the helix is now easily understood:
\bb
x^\perp(t)\sim|\eta|\,\sqrt{1-2\,\cos\,\frac{x^1(t)+1/a'(t)-1/a'_e}{|\eta|}\, \,[1-a'_e\,x^1(t)]+[1-a'_e\,x^1(t)]^2}.
\ee
It is bounded by
\bb |\eta|\,\Big|1-|1-a'_e\,x^1(t)|\,\Big|\le x^\perp(t)\le \,|\eta|\,\Big|1+|1-a'_e\,x^1(t)|\,\Big|.
\label{bounds}
\ee
These bounds are indicated in Figure \ref{xT}, where they touch at time $t_{\rm vanish}=0.66\, {\rm as}$ in our example of photons with redshift $z=2.4$. 
The comoving distance between the light-like geodesic and the position of the photon $x^\perp(t)$ may grow unboundedly as $|\eta|\,a'_e\,x^1(t)$. However for reasonable cosmologies, redshifts
and atomic periods, this physical distance $a(t)x^\perp(t)$ remains of the order of the wavelength.

The two initial helicity states $s=\pm\ \hbar$ yield identical helices except that they turn in opposite directions. Therefore today, the physical offset between the two positions at arrival oscillates between 0 and $(z+1)\,T/\pi$ with  period 
\bb
{\textstyle\frac{1}{2}}\, T_{\rm helix}(t_0)={\textstyle\frac{1}{2}} \,\frac{z+1}{1+q_0}\,T. \label{arrival}\ee

\section{Conclusion}

We have proposed to try and understand the effect of the photonic spin on the relativistic trajectories of light in the gravitational field of our expanding universe. We have found it worthwhile to start from the Mathisson-Papapetrou-Dixon equations derived by means of the principle of general covariance adapted to this context by Souriau \cite{Sou74}. The model has then been specialized to the massless case by means of the specific equation of state (\ref{SP=0}), leading to the Souriau-Saturnini equations (\ref{dotXter})--(\ref{dotSter}), a system of nine coupled first-order ordinary differential equations involving position, $\bx(t)$, spatial momentum, $\hbp(t)$, and spin vector, $\bs(t)$.

These equations of motion have two well-known delicate properties:
\begin{itemize}
\item
They are degenerate when the consistency condition $R(S)(S)\neq0$ (see (\ref{RSSneq0})) is violated which happens in particular in the fieldfree case. When the equations are not degenerate, one says that the (external) gravitational field localizes (spinning) photons. Already in 1976 Saturnini \cite{Sat76}  has proved that the Schwarzschild metric localizes photons and he has computed their trajectories. In the same setting it has been observed that an external electromagnetic field localizes chiral fermions  \cite{DH15,EDHZ16}; see also the recent contribution \cite{Sto15}. Let us also note that this approach yields a geometric derivation \cite{DHH1,DHH2} of the Fedorov-Imbert effect \cite{Fed55,Imb72} which has been confirmed experimentally \cite{BNKH08,HK08} in 2008.  
\end{itemize}
The present work proves that generic Robertson-Walker metrics do also localize photons.
\begin{itemize}
\item
When not degenerate the equations of motion involve tachyonic velocities, see Equation~(\ref{tachyon}). The discontinuity in the Fedorov-Imbert effect is an extreme manifestation of this property, as the velocity is infinite at the discontinuity.
\end{itemize}

The six conservation laws associated with the isometries coming from the cosmological principle, as well as the conserved ``energy'' coming from the conformal-Killing (temperature) vector field are worked out. Those are used to eliminate momentum and spin from the previous system of equations, to obtain the three differential equations for the trajectory. We compute it for two cases: for the flat Lambda Cold Dark Matter model numerically, and for generic flat Robertson-Walker models by linearizing the equations using two small parameters, namely the suitably normalized scalar spin, $s$, and the transverse spin, $s^\perp_e$, at emission of the photon.

Let us recall now that the photon remains our most faithful go-between on all distance scales; it makes no detours and travels untiringly at full throttle $c$. The photon is also a taciturn messenger; it only tells us its incoming direction and its spectrum. 

This picture is modified if we acknowledge that the photon carries spin. The spin gives an internal structure to the photon; its straight trajectory in the gravitational field of an expanding universe curls into two helices, one for each polarization, and we must correct our notion of full throttle. And maybe, if we listen carefully, a photon having travelled cosmic scales has more to tell us than just direction and spectrum.

The quantization rules of spin prohibit that the projection of spin on a given direction is equal to plus or minus the absolute value of spin. Therefore  spin cannot be parallel nor anti-parallel to its momentum, $\bs^\perp\not=0$, ``it cannot be enslaved''. 
Then by Equation~(\ref{tachyon}) the speed of the spinning photon exceeds $c$. Let us recall that there is another instance where we must admit particles with tachyonic velocities: the Feynman propagator of the Dirac operator has a support that leaks out of the light-cone~\cite{scharf}. However an exponential damping of the leakage prevents transmission of signals with tachyonic speed. In our situation, the tachyonic velocity of the photon and the form of the helix conspire and prevent transmission of signals with tachyonic speed over distances exceeding a few wavelengths of the photon, see Figure \ref{xL}.

Of course the main question is: can the periodically varying offset between the two polarization states of the photon at arrival be measured. Recall that this offset turns out to be of the order of the wavelength of the light just as for the Fedorov-Imbert effect. If the offset can be measured in the cosmological setting, Equation (\ref{arrival}) would give us a new independent measurement of the acceleration of the universe today.

\bigskip

\textbf{Acknowledgments:} It is a pleasure to thank P. Horv\'athy and A. Tilquin for their friendly advice. We are also indebted to our referee. His precise comments allowed us to improve the manuscript and to fill a gap in our proof in the appendix.


\renewcommand\thesection{A}
\numberwithin{equation}{section}

\section{Appendix: Proof of Eqs (\ref{P2const}), (\ref{dotXter}) \& (\ref{dotPter})}\label{AppendixA}

As a preparation, we mention the useful relationship \cite{Duv72,Sou74}
\begin{equation}
ABA=+\half\Tr(AB)A+\Pf(A)\star B
\label{magic}
\end{equation}
for all skewsymmetric linear operators $A,B$ of the tangent plane of $M$ at $X$.


Let us  start with the MPD equations, and assume the equation of state~(\ref{SP=0}) to hold.

1) One easily shows using (\ref{dotS}) that $d(\Tr(S^2))/d\tau=2\Tr(S\dS)=0$, leading to Eq.~(\ref{P2S2const}).

2) Our first task is to prove that $P^2$ is a constant of motion provided the consistency condition (\ref{CompatibilityCondition}) is everywhere satisfied.
Since $\dS{}P+S\dP=0$, we have
\begin{equation}
P(\barP\dX)-\dX{}P^2-\half{}SR(S)\dX=0
\label{Eq1}
\end{equation}
using (\ref{dotP}) and (\ref{dotS}), and also $\barP\dS\dP=0$ by the skewsymmetry of $S$. The latter equation and (\ref{dotSbis}) imply $\barP\dX\,\barP\dP=\dot{\barP}\dX\,P^2=0$ in view of (\ref{dotP}) and the skewsymmetry of $R(S)$. We conclude that
\begin{equation}
\barP\dX\,\frac{d P^2}{\!\!\!d\tau}=0
\label{Eq2}
\end{equation}
which duly entails that $P^2=\const$ (see (\ref{P2const})) or $\barP\dX=0$. It thus remains to investigate the latter possibility. Equation (\ref{Eq1}) reduces to $\dX{}P^2+\half{}SR(S)\dX=0$; multiplying by $R(S)$ from the left, we obtain $R(S)\dX{}P^2+\half{}R(S)SR(S)\dX=0$ and hence $\left[P^2+\frac{1}{4}\Tr(SR(S))\right]R(S)\dX+\half\Pf(R(S))(\star\,S)\dX=0$ by means of (\ref{magic}). Multiplying again by $S$ on the left, we end up with $\left[P^2+\frac{1}{4}\Tr(SR(S))\right]SR(S)\dX=0$ since $\Pf(S)=0$ in view of (\ref{SP=0}). Then, $SR(S)\dX=0$ provided 
\begin{equation}
\nu:=P^2+\frac{1}{4}\Tr(SR(S))\neq0
\label{nu}
\end{equation}
 which we suppose throughout, and thus (\ref{Eq1}) yields $P^2=0$ if $\dX\neq0$ everywhere (which is tacitly assumed). This proves~(\ref{P2const}) assuming  $\nu\neq0$, i.e., the consistency condition~(\ref{CompatibilityCondition}).
 
3) Let us consider, for completeness, the massive case where $P^2=m^2=\const>0$. Eq.~(\ref{Eq1}) implies that $\dX=\lambda P+SV$ for some scalar $\lambda$ and vector $V$ defined on the worldpath of the particle. Using again Eq. (\ref{Eq1}), we get $\nu{}SV=-\half\lambda{}SR(S)P$ where $\nu$ is as in (\ref{nu}) --- this is obtained by using (\ref{magic}) with $A=S$ \& $B=R(S)$, and $\Pf(S)=0$. To sum up, we find the following explicit expression of the velocity, namely
\begin{equation}
\dX=P-\frac{1}{2\nu}SR(S)P
\label{intermediatedX}
\end{equation}
with the above notation, and with the choice $\lambda=1$ enabled by reparametrization.

\goodbreak

4) Let us prove Eq. (\ref{dotXter}) in the massless case $P^2=0$, see (\ref{ZeroMass+Spin}). The  equation of state (\ref{SP=0}) implies that $S=\star(P\wedge{}Q)$ for some vector $Q$ not parallel to $P$. An easy computation shows that $-\frac{1}{2}\Tr(S^2)=(\barP Q)^2$, hence $|\barP{}Q|=|s|$, and $\barP{}Q\neq0$ since $s\neq0$.

Returning to (\ref{Eq1}), we find that $\barP\dX=0$ because $SQ=0$ and $\barP Q\neq0$. Indeed, since $S=\star(P\wedge{}Q)$, we necessarily have $SQ=0$. Now Equation (\ref{Eq1}) with $P^2=0$ reads $P (\barP\dX)-\frac{1}{2}S R(S)\dX=0$. Taking the scalar product of the latter by $Q$ yields $(\barQ{}P)(\barP\dX)=0$, hence $\barP\dX=0$ because $\barP{}Q\neq0$. Let us again decompose the velocity, this time in full generality, as $\dX=\lambda P+\mu Q+SV$ for some scalars $\lambda,\mu$ and a vector $V$. From $\barP\dX=\lambda P^2+\mu\barP{}Q+\barP{}SV=0$, we deduce that $\mu$ vanishes and that $\dX=\lambda P+SV$ just as before. The preceding computation leads therefore to the same form (\ref{intermediatedX}) of the velocity together with the expression (\ref{nu}) of $\nu$ with $P^2=0$, that is $\nu=-\frac{1}{4}R(S)(S)$ with Definition (\ref{DefRSS}). This ends the proof of (\ref{dotXter}).

Let us emphasize at this point that the velocity (\ref{intermediatedX}) of spinning massive particles depends continuously on the mass, $m$, and rather strikingly coincides with the sought velocity (\ref{dotXter}) in the limit $m\to0$, of course, provided the consistency condition (\ref{nu}) holds true.


5) To finish, we provide the proof of Eq. (\ref{dotPter}). We use (\ref{dotP}) and (\ref{intermediatedX}) to claim that 
\begin{eqnarray}
\nonumber
\dP&=&-\half{}R(S)\dX\\[4pt]
&=&
\nonumber
-\half\left[R(S)P-\frac{1}{2\nu}R(S)SR(S)P\right]\\[6pt]
&=&
-\frac{\Pf(R(S))}{R(S)(S)}\,\star(S)P
\end{eqnarray}
with the help of (\ref{magic}) with, this time, $A=R(S)$ \& $B=S$. 

Equation (\ref{dotPter}) follows by noting that the Pauli-Lubanski vector $W=\star(S)P$ is
\begin{equation}
W=s\,P
\label{Pauli-Lubanski}
\end{equation}
for massless particles of helicity $s$.



\begin{thebibliography}{99}

\bibitem{BI75}
I. Bailey, W. Israel,
``Lagrangian Dynamics of Spinning Particles and
Polarized Media in General Relativity'',
Commun. Math. Phys. {\bf 42} (1975) 65.
\url{http://projecteuclid.org/euclid.cmp/1103898967}

\bibitem{BB04}
K. Yu. Bliokh, Yu. P. Bliokh, 
``Topological spin transport of photons: the optical Magnus Effect and Berry Phase'', 
Phys. Lett. A {\bf 333}, 181--186, (2004).
\url{http://lanl.arxiv.org/pdf/physics/0402110v1},
\url{http://dx.doi.org/10.1016/j.physleta.2004.10.035}

\bibitem{BB06}
K. Yu. Bliokh, Yu. P. Bliokh, 
``Conservation of Angular Momentum, Transverse Shift, and Spin Hall Effect in Reflection and Refraction of Electromagnetic Wave Packet'', 
Phys. Rev. Lett. {\bf 96} (2006) 073903.
\url{http://lanl.arxiv.org/abs/physics/0508093},
\url{http://dx.doi.org/10.1103/PhysRevLett.96.073903}

\bibitem{BNKH08}
K. Yu. Bliokh, A. Niv, V. Kleiner, E. Hasman, 
``Geometrodynamics of Spinning Light'',
Nature Photon. {\bf 2} (2008) 748.
\url{http://lanl.arxiv.org/abs/0810.2136},
\url{http://dx.doi.org/10.1103/PhysRevLett.96.073903}

\bibitem{BA13}
K. Yu. Bliokh, A. Aiello,
``Goos-H\"anchen and Imbert-Fedorov beam shifts: An overview'',
J. Opt. {\bf 15} (2013) 014001 (16pp).
\url{http://lanl.arxiv.org/abs/1210.8236},
\url{http://dx.doi.org/10.1088/2040-8978/15/1/014001}

\bibitem{BN15}
K. Yu. Bliokh, F. Nori,
``Transverse and longitudinal angular momentum of light'',
Phys. Rept. {\bf 592} (2015) 1--38.
\url{http://lanl.arxiv.org/abs/1504.03113},
\url{http://dx.doi.org/10.1016/j.physrep.2015.06.003}

\bibitem{Dix70}
W. G. Dixon,
``Dynamics of Extended Bodies in General Relativity. I. Momentum and Angular Momentum'',
Proc. R. Soc. Lond. A {\bf 314} (1970). 
\url{http://rspa.royalsocietypublishing.org/content/314/1519/499}

\bibitem{DFS72}
C. Duval, H. H. Fliche, J.-M. Souriau,
``Un mod\`ele de particule \`a spin dans le champ gravitationnel et \'electromagn\'etique'',
CRAS, {\bf 274} S\'erie A (1972) 1082.
\url{http://gallica.bnf.fr/ark:/12148/bpt6k56190683/f60.image.r=Souriau?rk=42918;4}

\bibitem{Duv72}
C. Duval,
``Un mod\`ele de particule \`a spin dans un champ \'electro\-magn\'etique
et gravitationnel ext\'erieur'',
th\`ese de 3\`eme cycle en physique th\'eorique, Universit\'e de Provence
(1972), unpublished.

\bibitem{Duv08}
C. Duval,
``Finsler Spinoptics'',
Commun. Math. Phys. {\bf 283} (2008) 701.
\url{http://arxiv.org/abs/0707.0200}

\bibitem{Duv12}
C. Duval,
``Polarized Spinoptics and Symplectic Physics'',
\url{http://xxx.lanl.gov/abs/1312.4486}.

\bibitem{DF78}
C. Duval, H. H. Fliche,
``A conformal invariant model of localized spinning test particles'',
J. Math. Phys. {\bf 19} (1978) 749.
\url{http://dx.doi.org/10.1063/1.523729}


\bibitem{DHH1}
C.~Duval, Z.~Horv\'ath, P.~Horv\'athy,
``Geometrical Spinoptics and the Optical Hall Effect'',
J. Geom. Phys {\bf 57} (2007) 925.
\url{http://arxiv.org/abs/math-ph/0509031},
\url{http://dx.doi.org/10.1016/j.geomphys.2006.07.003}

\bibitem{DHH2}
C.~Duval, Z.~Horv\'ath, P.~Horv\'athy,
``Fermat Principle for spinning light'',
Phys Rev D
{\bf 74} (2006) 021701(R).
\url{http://arxiv.org/abs/cond-mat/0509636},
\url{http://dx.doi.org/10.1103/PhysRevD.74.021701}

\bibitem{DH15}
C. Duval, P.A. Horv\'athy
``Chiral fermions as classical massless spinning particles'',
Phys. Rev. D {\bf 91} (2015) 045013.
\url{http://dx.doi.org/10.1103/PhysRevD.91.045013},
\url{http://arxiv.org/abs/1406.0718v4}

\bibitem{EDHZ16}
M. Elbistan, C. Duval, P. A. Horv\'athy, P.-M. Zhang,
``Helicity of spin-extended chiral particles'',
Physics Letters A {\bf 380} (2016) 1677.
\url{http://arxiv.org/abs/1508.02188},
\url{http://dx.doi.org/10.1016/j.physleta.2016.03.016}

\bibitem{Fed55}
F. I. Fedorov,
``To the theory of total reflection'',
Doklady Akademii Nauk SSSR
Vol. 105, \# 3 (1955) 465.
\url{http://master.basnet.by/congress2011/symposium/spbi.pdf}

\bibitem{FS11}
V. P. Frolov, A. A. Shoom,
``Spinoptics in a stationary spacetime'',
Phys. Rev. D {\bf 84} (2011) 044026.
\url{http://lanl.arxiv.org/abs/1105.5629},
\url{http://dx.doi.org/10.1103/PhysRevD.84.044026}

\bibitem{GBM07}
P. Gosselin, A. B\'erard, H. Mohrbach,
``Spin Hall effect of Photons in a Static Gravitational Field'',
Phys. Rev. D {\bf 75} (2007) 084035.
\url{http://arxiv.org/abs/hep-th/0603227},
\url{http://dx.doi.org/10.1103/PhysRevD.75.084035}

\bibitem{HLGBN}
E. Harms, G. Lukes-Gerakopoulos, S. Bernuzzi, A. Nagar,
`Spinning test-body orbiting around Schwarzschild black hole: circular dynamics and gravitational-wave fluxes''.
\url{https://arxiv.org/abs/1609.00356}

\bibitem{HK08}
O. Hosten, P. Kwiat, 
``Observation of the Spin Hall Effect of Light via Weak Measure\-ments'',
Science {\bf 319}: 5864 (2008) 787--790.
\url{http://dx.doi.org/10.1126/science.1152697}

\bibitem{Imb72}
C. Imbert,
``Calculation and Experimental Proof of the Transverse Shift Induced by Total Internal Reflection of a Circularly Polarized Light Beam'',
Phys. Rev. D {\bf 5} (1972) 787.
\url{http://dx.doi.org/10.1103/PhysRevD.5.787}

\bibitem{Kun72}
H. P. K\"unzle, 
``Canonical dynamics of spinning particles in gravitational and
electromagnetic fields'', 
J. Math. Phys. {\bf 13} (1972) 739. 
\url{http://dx.doi.org/10.1063/1.1666045}

\bibitem{Mat37}
M. Mathisson,
``Neue Mechanik materieller Systeme''
Acta Phys. Pol. {\bf 6} (1937) 163; ``Das zitternde Elektron und seine Dynamik'', Acta Phys. Pol. {\bf 6} (1937) 218.

\bibitem{pdg}
K.A. Olive et al. (Particle Data Group), ``Review of particle Physics'',
Chinese Physics C {\bf  38} (2014) 090001 1.
\url{http://dx.doi.org/10.1088/1674-1137/38/9/090001}

\bibitem{OMN04}
M. Onoda, S. Murakami, N. Nagaosa, 
``Hall Effect of Light'', 
Phys. Rev. Lett. {\bf 93}, 083901, (2004).
\url{http://lanl.arxiv.org/abs/cond-mat/0405129},
\url{http://dx.doi.org/10.1103/PhysRevLett.93.083901}

\bibitem{Papa51}
A. Papapetrou, 
``Spinning Test-Particles in General Relativity. I'',
Proc. Roy. Soc. A {\bf 209} (1951) 248.
\url{http://rspa.royalsocietypublishing.org/content/209/1097/248}

\bibitem{Sat76}
P. Saturnini,
``Un mod\`ele de particules \`a spin de masse nulle dans le champ de gravitation'',
Th\`ese de 3\`eme cycle en physique th\'eorique, Universit\'e de Provence (1976). \url{https://hal.archives-ouvertes.fr/tel-01344863v1}

\bibitem{scharf}
G.~Scharf,
  \textsl{Finite quantum electrodynamics},
  Berlin, Springer (1989), Chapter 2.3.

\bibitem{SSD}
J.-M. Souriau,
\textsl{Structure des syst\`emes dynamiques}, Dunod (1970, \copyright\,1969);
\textsl{Structure of Dynamical Systems. A Symplectic View of Physics},
Birkh\"auser, 1997.

\bibitem{Sou74}
J.-M. Souriau,
``Mod\`ele de particule \`a spin dans le champ \'electro\-magn\'etique et gravitationnel'',
Ann. Inst. Henri Poincar\'e \textbf{20 A} (1974), 315.
\url{http://www.jmsouriau.com/Publications/JMSouriau-ModPartSpin1974.pdf}

\bibitem{Sou75}
J.-M. Souriau,
\textit{M\'ecanique statistique, groupes de Lie et cosmologie}, in G\'eom\'etrie symplectique et physique math\'ematique (Colloq. Internat. CNRS No. \textbf{237}, Aix en Provence, 1974), pp. 59--113. Editions Centre Nat. Recherche Sci., Paris, 1975.

\bibitem{Ste78}
 S. Sternberg,
\textit{On The Role Of Field Theories In Our Physical Conception Of Geometry},
 in: Bonn 1977, Proceedings, Differential Geometrical Methods in Mathematical Physics II, Lecture Notes in Mathematics \textbf{676}, Berlin 1978, pp. 1--80.
 
\bibitem{Sto15}
  M.~Stone,
``Berry phase and anomalous velocity of Weyl fermions and Maxwell photons'',
  Int. J. Mod. Phys. B {\bf 30} (2016) 1550249.
 \url{http://dx.doi.org/10.1142/S0217979215502495},
\url{http://arxiv.org/abs/1507.01807}

\bibitem{Tau64}
A. H. Taub,
``Motion of Test Bodies in General Relativity'',
J. Math. Phys. {\bf 5} (1964) 112.

\bibitem{TB15}
K. S. Thorne, R. D. Blandford,
\textsl{Modern Classical Physics: Optics, Fluids, Plasmas, Elasticity, Relativity, and Statistical Physics}, Princeton University Press (2017).
\url{http://www.pmaweb.caltech.edu/Courses/ph136/yr2012/}

\bibitem{Tul59}
W. Tulczyjew, 
``Motion of multipole particles in general relativity theory'', 
Acta Phys. Pol. {\bf 18} 
(1959) 393.

\bibitem{ZHA14}
N. Zalaquett, S. A. Hojman,  F. A. Asenjo,
``Spinning massive test particles in
cosmological and general static spherically
symmetric spacetimes'',
Class. Quantum Grav. {\bf 31} (2014) 085011.
\url{http://dx.doi.org/10.1088/0264-9381/31/8/085011},
\url{http://lanl.arxiv.org/abs/1308.4435}


\end{thebibliography}
\end{document}